# Efficient and environmental-friendly perovskite solar cells via embedding plasmonic nanoparticles: an optical simulation study on realistic device architecture

George Perrakis,*,1,2 George Kakavelakis,*,3,4 George Kenanakis,1 Constantinos Petridis,3,5 Emmanuel Stratakis,1 Maria Kafesaki,1,2 and Emmanuel Kymakis3

[1]*Institute of Electronic Structure and Laser (IESL), Foundation for Research and Technology Hellas (FORTH), Heraklion, Greece*
[2]*Dept. of Materials Science and Technology, Univ. of Crete, Heraklion, Greece*
[3]*Center of Materials Technology and Photonics and Electrical Engineering Department, Technological Educational Institute (TEI) of Crete, Heraklion, Crete, 71004, Greece*
[4]*Cambridge Graphene Centre, University of Cambridge, 9 JJ Thomson Avenue, Cambridge CB3 0FA, UK*
[5]*Department of Electronic Engineering Technological Educational Institute (TEI) of Crete, Chania 73132, Greece*
*Corresponding authors:*
*\*E-mail: gperrakis@iesl.forth.gr*
*\*E-mail: gk415@cam.ac.uk*

**Abstract:** Solution-processed, lead halide-based perovskite solar cells have overcome important challenges over the recent years, offering low-cost and high solar power conversion efficiencies. However, they still undergo unoptimized light collection due mainly to the thin (~350 nm) polycrystalline absorber layers. Moreover, their high toxicity (due to the presence of lead in the perovskite crystalline structure) makes it necessary that the thickness of the absorber layers to be further reduced, for their future commercialization, without reducing the device performance. Here we aim to address these issues via embedding spherical plasmonic nanoparticles of various sizes, composition, concentrations, and vertical positions, for the first time in realistic halide-based perovskite solar cells architecture, and to clarify their effect on the absorption properties and enhancement. We theoretically show that plasmon-enhanced near-field effects and scattering leads to a device photocurrent enhancement of up to ~7.3% when silver spheres are embedded inside the perovskite layer. Interestingly, the combination of silver spheres in perovskite and aluminum spheres inside the hole transporting layer (PEDOT:PSS) of the solar cell leads to an even further enhancement, of up to ~12%. This approach allows the employment of much thinner perovskite layers in PSCs (up to 150 nm) to reach the same photocurrent as the nanoparticles-free device and reducing thus significantly the toxicity of the device. Providing the requirements related to the size, shape, position, composition, and concentration of nanoparticles for the PSCs photocurrent enhancement, our study establishes guidelines for a future development of highly-efficient, environmentally friendly and low-cost plasmonic perovskite solar cells.

## 1. Introduction

Emerging halide Perovskite ($CH_3NH_3PbX_3$, X= Cl, Br and I) based thin-film Solar Cells (PSCs) have attracted significant interest over the recent years due to their remarkable photovoltaic performance (23.7%). Some of their main characteristics are the high absorption coefficient of perovskite along with its direct band-gap property[1], high power conversion efficiency[2],[3] (PCE) combined with fabrication simplicity using solution-processing techniques at room temperature[4],[5]. However, despite their high performance, there are still two major issues that must be addressed. Firstly, a further improvement of the light collection of the PSCs should be achieved[6],[7]. Additionally, a reduction of the PSCs' toxicity[8], due to the presence of pure lead (Pb) in the perovskite materials, should be attained for their further commercialization. A possible solution to reduce the amount of the lead is to employ perovskite absorbers thinner than the optimum thickness[9] of about ~350 nm without though deteriorating the absorption of the solar cell. However, there is a main drawback using this approach, which is the small interaction time of the incoming wave with the very thin perovskite layer resulting to an unoptimized light collection.

One of the most promising approaches to increase the light/matter interaction time and as a result to improve light collection in thin-film solar cells is the use of the plasmonic effect[10]. Surface plasmons are collective oscillations of conduction electrons of metallic nanoparticles that are excited by light at the nanoparticle interface with the surrounding dielectric medium. Important features of the surface plasmons are that they are associated with high local-field amplitudes and strong far-field scattering at the resonances of the oscillations. Strong far-field scattering can increase the absorption efficiency in thin film solar cells by exploiting the effect of total internal reflection. This has been already demonstrated in amorphous-silicon-based thin-film solar cells[11] and in organic solar cells[12]. Moreover, the high local fields in the vicinity of plasmonic nanoparticles overlap not only with metal but also with the surrounding absorbing matter resulting in increased absorption, A. This follows directly from Poynting's theorem for power dissipation[13] [$A \sim |E|^2$]. To benefit from the above features, the parameters determining the resonance characteristics of the plasmonic nanoparticles, i.e. the size, position or the material of the nanoparticles and their hosting environment[14], must be carefully chosen to properly match the plasmon resonance to the spectral properties of the solar cell's material. Otherwise, ohmic losses or coupling between the nanoparticles[15] can lead to increased parasitic absorption in metal turning into heat, a behavior not desirable in solar cells.

Already, several studies have explored the plasmonic enhancement in perovskite thin-films. An additional reason for this is that metal particles dispersed inside a solution preserve the simplicity of the PSC at processing without increasing its cost[16]. The main finding of the existing studies was that metal particles could improve considerably the beneficial absorption inside the perovskite material[17]. Interestingly, parasitic absorption in metal particles does not surpass the enhancement they provide. More specifically, both N.K. Pathak et al.[18] and Roopak et al.[19] explored Mie theory[20] and showed that silver, gold and aluminum nanoparticles, embedded inside a perovskite matrix, support plasmonic resonances of high magnitude with tunable resonance frequency and width by varying the size, shape or material. These results are very promising since plasmonic nanoparticles can induce light incoupling to the perovskite near its band edge (~650-800 nm) where it shows inferior photo response due to lower absorption in conjunction with increased reflection from the perovskite at that regime[6],[7]. Moreover, Carretero-Palacios S. et al.[21],[22] theoretically exploited the effect of plasmonic enhancement by incorporating metallic nanoparticles of different shapes, sizes, concentrations and composition in methyl ammonium lead iodide perovskite ($CH_3NH_3PbI_3$) absorbing layers assumed to be supported on a glass substrate and coated by a hole transporting material. Notably, it was shown that the conditions for plasmonic enhancement are not very stringent and ample ranges of sizes, shapes and concentrations of the metallic nanoparticles give rise to an important degree of improvement. In addition, the main conclusion was that the improvement was neatly the result of the near optical field enhancement at longer wavelengths within the absorption band of perovskites (~550-800 nm).

These studies demonstrate the possibility of improvement of perovskite film absorption with the employment of metallic nanoparticles; it is still of great importance, though, if they can be successfully implemented to realistic PSCs. In a real solar cell one has to consider the interplay of metal particles with several optical mechanisms; for instance the effect of the asymmetric environment of the PSC on the localized surface plasmon resonances of the nanoparticles compared to a homogeneous perovskite matrix, coupling with fabry-perrot (F.P.)

resonances (i.e. modes that are supported by the absorbing layer due to its finite thickness), interference effects and reflections introduced by the multilayered structure of the PSC, etc., which give rise to a more complex optical system in the neighborhood of the nanoparticles. All those effects are highly unexplored and need careful examination as to definitively conclude on the impact of plasmonic nanoparticles on the performance of realistic PSCs and to be able to optimize this impact.

In this respect we aim in this paper to clarify the effect of the plasmonic nanoparticles on the absorption properties and enhancement of realistic PSCs, with further aim to improve the optical performance of such PSCs. We achieve this by incorporating metal nanoparticles of different size, concentration and material (among the most standard ones, namely silver, gold and aluminum) at different positions into the perovskite absorber or inside different layers of the PSC (see Fig. 1). Moreover, the combined effect of nanoparticles embedded inside different layers at the same time is examined. The aim is to (i) achieve an enhanced absorption efficiency compared to the conventional (nanoparticles-free) PSCs and (ii) reduce the amount of lead by either replacing perovskite material by nanoparticles or by employing thinner perovskite absorbers, without deteriorating the absorption. The perovskite material that we deal with in this work is the most common halide perovskite, the methyl ammonium lead iodide perovskite ($CH_3NH_3PbI_3$), which has been extensively examined for solar cells. One of its main advantages is the fabrication simplicity, while its direct band-gap of ~1.55 eV (~800 nm, i.e. at the onset of the optical range), very close to the ideal compared to other perovskites in which another halide is present, leads to high efficiencies. Improving further its already good performance without increasing the fabrication cost is of high importance.

## 2. PSC geometrical and material parameters, and modeling approaches

Here we investigate the common inverted[23] planar heterojunction PSC geometry (see Fig. 1) where the poly(3,4-ethylenedioxythiophene) polystyrene sulfonate (PEDOT:PSS) and the phenyl-C71-butyric acid methyl ester (PCBM) serve as the hole transporting layer (HTL) and the electron transporting layer (ETL) respectively. The planar inverted (upside down fabrication, possibly on top of flexible substrates[23]) PSC adopts the structure of organic solar cells to fulfill the requirements of high performance as well as low-cost and easy fabrication where all the device layers could be deposited at room temperature through a solution-process[4]. Additionally, since the most efficient two-terminal monolithic Si/Perovskite tandem solar cells are based on inverted heterojunction device architecture, its fine optimization is of significant importance towards the beyond 30% efficiency target. The structure of the device is as follows: $SiO_2$ (1.1 mm) /ITO (100 nm) /PEDOT:PSS (40 nm) /$CH_3NH_3PbI_3$ /PCBM (50 nm) /Al (100 nm) (see Fig. 1), where the numbers indicate the thickness of each layer. The transparent conducting oxide (ITO: indium tin oxide) and the aluminum back-reflector serve as the electrical contacts.

In our study the thickness of the perovskite absorber is varied among the standard lengths of PSCs, regarding efficient absorption and photocarrier collection[9], from 150 nm to 400 nm. Moreover, the refractive index and the extinction coefficient of the perovskite absorber [see Fig. S1(a)] and the other layers are obtained from Kakavelakis et al.[4], and those of silver, gold and aluminum from Palik[24]. Despite the "inverted" fabrication sequence of the PSC layers, we will refer as top layers the layers at which the sun is incident on, i.e., following this order from the top to bottom: Glass/ITO/PEDOT:PSS/Perovskite/PCBM/Al.

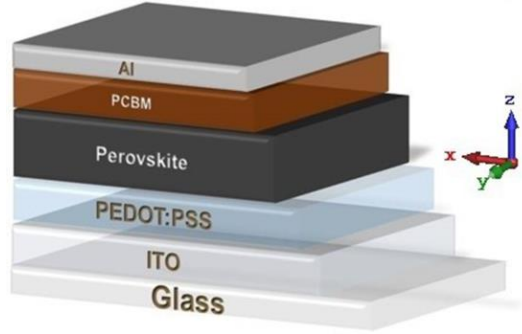

Fig. 1. Geometry of the inverted planar heterojunction perovskite solar cell. The thickness and role of the different layers of the cell are discussed in the main text.

Due to the complexity of the solar cell geometries investigated here, which include metallic nanoparticles and highly absorbing multilayered materials, a numerical methodology is required. To evaluate the optical performance of the plasmonic PSCs, we perform three-dimensional full-wave electromagnetic simulations using the commercially available software CST Microwave Studio (Computer Simulation Technology GmbH, Darmstadt, Germany) based on the finite element method. The simulated structure consists of a semi-infinite $SiO_2$ (glass) slab, which serves as a substrate for the device fabrication. The simulated region is terminated by unit cell (Floquet) boundary conditions at the planes perpendicular to the surface of the PSC, to investigate the effect of periodically placed nanoparticles, and by open-boundary conditions at the planes parallel to the PSCs' surfaces.

Finally, we quantify the optical performance of the device with and without the metallic nanoparticles by obtaining numerically the absorption in the perovskite material from which we determine the generated current density inside the solar cell, called the illumination or photocurrent density ($J_{ph}$), given by:

$$J_{ph} = q \int A_p(\lambda) \Phi_{AM1.5G}(\lambda) d\lambda, \qquad (1)$$

In Eq. (1) $\Phi_{AM1.5G}$ is the photon flux density [in photons·m$^{-2}$·s$^{-1}$·nm$^{-1}$] of the "AM 1.5G" standard sunlight spectrum[25] reaching the Earth's surface, that is considered universal when characterizing solar cells, and $A_p$, $q$ are the absorption of the perovskite material and the elementary charge [in C] of an electron respectively. The integration takes place at 300<$\lambda$<800 nm that corresponds to the range within the pass-band edges of the perovskite material. Throughout this work, $J_{ph}$ is considered the most critical parameter for the performance evaluation of the PSCs. Increasing $J_{ph}$ leads to an increased current density under short circuit conditions ($J_{SC}$) and hence increased extracted electrical power and efficiency, as long as the approach chosen for the $J_{ph}$ enhancement does not have a negative influence on the electrical properties of the device, for example due to a possible increase of the interface transport resistance arising by the Schottky barrier at metal-semiconductor junctions.

## 3. Results and discussion

In the current section we discuss the effect on the PSC photocurrent density ($J_{ph}$) of plasmonic nanoparticles of spherical shape and of different sizes, concentrations and materials embedded in different parts of the PSC, for the cell shown in Fig. 1. To gain insight on the mechanisms that lead to the unexploited absorption losses which decline the absorption efficiency of the solar cell and to employ a common reference system we investigate and discuss first the pristine PSC, i.e. the cell with no metallic nanoparticles (see subsection 3.1). Next, we discuss the effect of plasmonic spheres inside the perovskite layer of the PSC (subsection 3.2) and demonstrate and analyze the associated increase of photocurrent density. In subsection 3.3 we examine the effect of plasmonic nanoparticles (not only spherical but also cylindrical ones) if placed inside the carrier transporting layers and not the perovskite layer. This way there is no competition between the volume occupied by the metal particles and the perovskite material. Finally, in subsection 3.4 the combination of nanoparticles placed inside more than one layers of the PSC is examined to demonstrate the full potential of plasmonic nanoparticles if combined with realistic PSCs.

## 3.1 Pristine PSC

To fully understand the PSC optical response and set appropriate references, we firstly calculate the absorption for each layer of the pristine PSC for a device as shown in Fig. 1 with perovskite layer thickness equal to 350 nm. The result is shown in Fig. 2(a). As can be seen, absorption in perovskite remains high along the entire optical spectrum, due to its high absorption coefficient [see Fig. S1(b)]. Interestingly, high absorption in perovskite persists even at longer wavelengths (~750 nm), close to its band-gap edge (~750-800 nm), as a result of its direct band-gap property. Specifically, perovskite absorbs 75% of the maximum achievable current density ($J_{ph,\ sun}$, obtained from Eq. (1) with $A_p=1$, integrated at $300<\lambda<800$ nm), a percentage which will be referred from this point as solar absorption, $(J_{ph}/J_{ph,\ sun})\cdot\%$, while all the other materials of the device absorb 8% (absorption in PCBM is almost zero along the entire spectrum). The highest fraction of the unexploited sun spectrum is lost due to reflection and equals 17%.

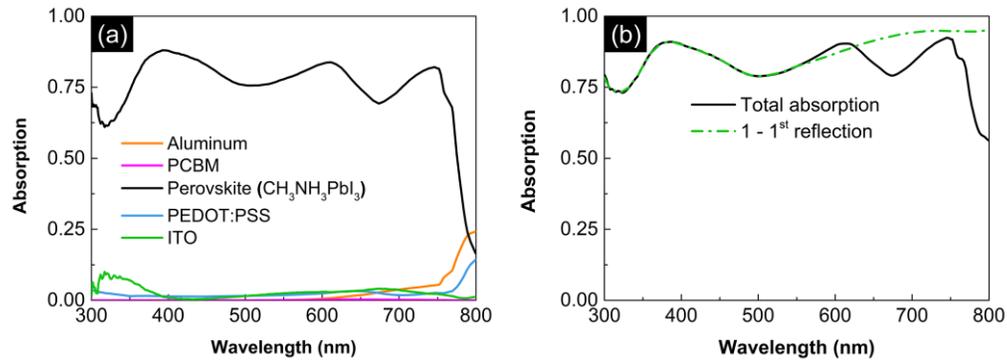

Fig. 2. (a) Absorption in each layer of the PSC. (b) Total absorption (black solid line) of the PSC, absorption in perovskite due to the first reflection only (green dashed-dotted line).

To reveal the origin of the reflection losses we calculated both the total absorption of the pristine PSC and the absorption $[A^{1st} = 1 - R^{1st}]$ arising from the "first" reflection ($R_{1st}$) of the top layers (ITO/PEDOT:PSS/Perovskite, assuming a semi-infinite perovskite layer) of the structure. The result is shown in Fig. 2(b). As can be seen, the total absorption of the pristine PSC and consequently of the perovskite, despite being high enough, is not optimized and there is a room for significant improvement. Moreover, the absorption spectrum of the system is divided into two spectral regimes of different behavior. For $\lambda<550$ nm, the total absorption of the pristine PSC follows the absorption arising from the "first" reflection (i.e. it occurs in the first pass of the wave into the perovskite) due to the very high absorption coefficient of the perovskite material [see Fig. S1(b)] for the corresponding wavelengths. For $\lambda>550$ nm, where the absorption coefficient of the perovskite material is much lower, interference effects (see Fig. S2) introduced by the several layers of the device of finite thickness are now present leading to an overall reduced absorption due to higher reflection losses relative to the "first" reflection. These results confirm that further improvement of the PSC could be achieved in agreement with other studies[6],[7], especially at the range of ~650-800 nm where perovskite material shows inferior absorption that results to higher reflection losses from the PSC. Such an improvement can be achieved by adoption of efficient light trapping strategies, which will forbid light from escaping. As we show in this article such a strategy is the proper incorporation of metallic nanoparticles.

Before proceeding to the incorporation of the nanoparticles we further evaluate our pristine PSC calculations by comparing with experimental data for the same device[16]. The $J_{ph}$ of our simulated pristine device was calculated at 20.40 mA/cm$^2$, very close to the experimentally obtained current density under short circuit conditions ($J_{SC}$) equal to 20.55 mA/cm$^2$. These results indicate that the experimental device achieves near-unity quantum yield for the generation and collection of charge carriers[26]. In addition, any mismatch between the simulated sunlight and the AM1.5G standard is negligibly small. The excellent agreement of our theoretical calculations with the experimental measurements allows us to continue with the examination of the plasmonic PSCs while using the $J_{ph}$ of the pristine PSC as a reference.

*3.2 Plasmonic nanoparticles inside perovskite*

In what follows, we aim to investigate and clarify the effect of plasmonic nanoparticles on the optical response of the PSC for different nanoparticle parameters and configurations. This will allow an efficient understanding and optimization of the PSC performance. In particular, we examine the effect of metal spheres embedded inside the perovskite layer for different metal-material (Subsection 3.2.1), sphere size and position (Subsection 3.2.2), and sphere concentration (i.e., occupied volume relative to the perovskite volume - Subsection 3.2.3); in all those studies we assume a constant perovskite layer thickness of 350 nm that is considered optimum regarding the PCE of the PSCs[9]. The spherical shape of nanoparticles was preferred when they are embedded inside the perovskite layer because it is associated with: (i) broadband extinction (scattering and absorption) cross-sections of high magnitude [see Fig. 3(a)], according to Mie theory[18], for an ample range of diameters, along the entire spectral region near perovskite's band edge (~650-800 nm) where it shows inferior photoresponse; (ii) significant field intensity enhancement in the vicinity of the nanoparticle[19]; (iii) preservation of the PSC fabrication simplicity due to its symmetric nature[27].

3.2.1 PSCs efficiency versus nanoparticle material

Optimization of plasmonic light trapping in solar cells is a balancing act in which several physical parameters must be taken into account. For instance, very small particles suffer from significant ohmic losses according to Mie theory[20] whereas larger particles come at the expense of the hosting material if their spacing (i.e. periodicity) is kept constant. To start our investigation from a nearly optimized size and periodicity setup, we performed first an optimization process (see in Fig. S3 optimization process for silver nanoparticles), where we required maximum photocurrent density enhancement, $\eta = (J_{ph,\ plasm} - J_{ph,\ ref})/J_{ph,\ ref}$, for the plasmonic PSC relative to the pristine PSC, for spheres located exactly at the middle of the perovskite layer (along z-direction – see Fig. 1) by changing the radius, r, of the spheres, and by changing the periodicity, $L$ ($L = L_x = L_y$), at which spheres are placed along the *x*- and *y*-axis.

Fig. 3(b) illustrates the absorption inside the perovskite for plasmonic PSCs that contain silver, gold and aluminum spherical particles located exactly at the middle of the perovskite layer with the size and periodicity for which the photocurrent density was found maximum following the previously mentioned optimization process (see Fig. S3). The associated to each case maximum photocurrent density along with the corresponding optimum radius and lattice constant are listed in Table 1.

As can be concluded from both Fig. 3(b) and Table 1, the highest absorption and thus photocurrent enhancement is obtained for silver nanoparticles. Moreover, the absorption enhancement for the case of silver and gold nanoparticles takes place at a wider spectral range compared to aluminum and coincides with the 2$^{nd}$ spectral region ($\lambda$>550 nm) mentioned in Section 3.1, where the absorption coefficient of the perovskite is lower.

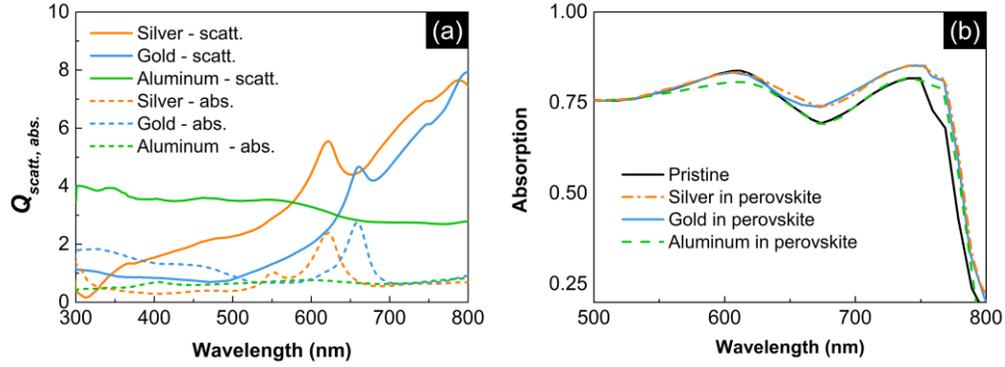

Fig. 3. (a) Scattering (solid lines) and absorption (dashed lines) cross-sections ($Q_{scatt., abs}$) of a silver sphere (orange line – sphere radius 40 nm), gold sphere (blue line- sphere radius 40 nm) and aluminum sphere (green line - sphere radius 60 nm) inside a homogeneous perovskite matrix (here the matrix is considered with no losses and the absorption cross-section represents only ohmic losses inside the spheres); (b) Absorption in perovskite for plasmonic PSCs with silver (orange dashed-dotted line), gold (blue line) and aluminum (green dashed line) spheres placed at the middle of the perovskite layer compared to the pristine case (black line).

**Table 1. Effect of nanoparticle material on the photocurrent density, $J_{ph}$, of a PSC with embedded spherical nanoparticles[a]) in the middle vertical position of the perovskite layer.**

| Nanoparticle material | $J_{ph}$ (mA/cm$^2$) | Enhancement factor, $\eta$ (%) | Optimum $r$ (nm) | Optimum $L$ (nm) |
|---|---|---|---|---|
| No nanoparticles | 20.40 | - | - | - |
| Silver | 21.07 | 3.24 | 40 | 300 |
| Gold | 20.98 | 2.98 | 40 | 300 |
| Aluminum | 20.46 | 0.25 | 60 | 300 |

[a]Spherical nanoparticles with optimized radius, $r$, and lattice constant, $L$, as to maximize the enhancement of $J_{ph}$.

In all cases, the origin of the absorption enhancement was that the plasmonic nanoparticles behaved as light nano-antennas that led to strong scattering (thus to increase of the interaction time of the field with the perovskite material) and to strong local fields (common to plasmonic antennas) in their vicinity, inside the absorptive perovskite layer. These strong local fields are demonstrated in Fig. 4, where the normalized distribution of the squared amplitude of the electric field, $|E|^2/|E_0|^2$, is plotted for an incident plane wave of $\lambda$=673 nm, at which the maximum absorption takes place ($E_0$ is the incident electric field). Fig. 4 demonstrates the higher local fields around the sphere for the case of silver and gold nanoparticles compared to the aluminum case. Moreover, from Fig. 3(a) one can observe the stronger scattered fields in the case of silver and gold compared to aluminum, especially at the region of $\lambda$>550 nm where the perovskite shows inferior absorption. Both high local and scattered fields make the silver nanoparticles the optimum plasmonic material choice for the solar cell absorption enhancement, justifying the corresponding larger enhancement factor (3.24%) mentioned in Table 1.

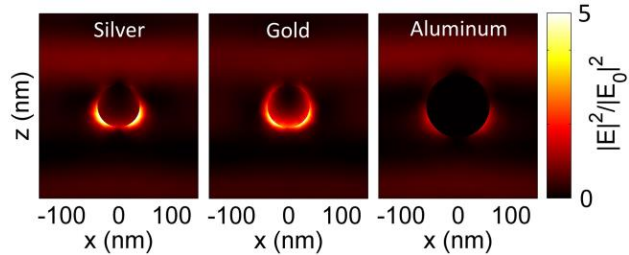

Fig. 4. Normalized (relative to the incident field) distribution of the squared amplitude of the electric field, at $\lambda$=673nm for silver, gold and aluminum (left, central, right figures respectively) spheres with optimum radius and periodicity for each case.

### 3.2.2. Vertical position and nanoparticle size effect on the PSCs performance

Another important parameter affecting the optical response of the PSC is the depth at which the nanoparticles are embedded inside the perovskite layer. In a homogeneous perovskite absorbing matrix the effect of nanoparticles' vertical position is related basically to the absorption depth of the wave into the perovskite[21]. For a more complex environment like in PSCs, the effect of nanoparticles' position is not so straightforward and further analysis is needed. Therefore, in this section, we vary the vertical position of silver spheres of the optimum case mentioned in Table 1 ($r$=40 nm, $L$=300 nm, parameters for which we found maximum solar absorption enhancement). Specifically, nanoparticles placed at four different vertical positions, namely at the top, middle, bottom and at perovskite bottoms (see inset in Fig. 5 - these points were considered sufficient for our analysis) were examined regarding the associated PSC absorption and resulting photocurrent density.

Fig. 5 shows the absorption inside the perovskite layer of the PSC device for all four vertical positions considered in comparison with the pristine PSC. The maximum absorption enhancement inside the perovskite for the "top" and the "middle" cases occurs at the proximity of $\lambda$=673 nm and is responsible for the higher solar absorption (equal to 77.5% and 77.3% respectively) relative to that of the "bottom" (75 %) and the "bottoms" case (75.6%) for which no substantial enhancement was found (for pristine PSC ~75%).

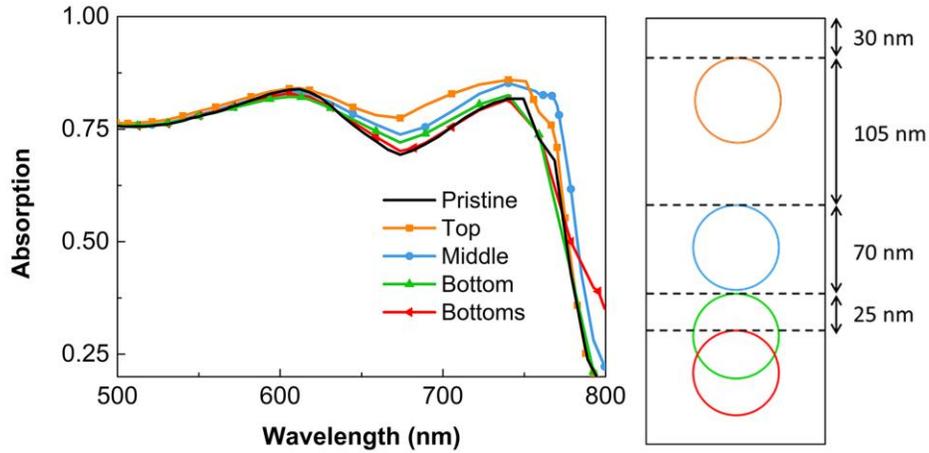

Fig. 5. Absorption of perovskite of the plasmonic PSC with silver spheres placed at different positions (top – orange line, middle – blue line, bottom – green line, bottoms – red line) inside the perovskite layer as depicted at the right inset, compared to the pristine case (black line).

Table 2. Effect on photocurrent density ($J_{ph}$) and efficiency enhancement for PSCs with silver spheres of different vertical position and different radius inside the perovskite layer[a].

| Vertical Position | $J_{ph}$ (mA/cm$^2$) | Enhancement factor, $\eta$ (%) | $r$ (nm) |
|---|---|---|---|
| No nanoparticles | 20.40 | - | - |
| Top | 21.12 | 3.50 | 40 |
| Middle | 21.07 | 3.24 | 40 |
| Bottom | 20.44 | 0.16 | 40 |
| Bottoms | 20.60 | 1.00 | 40 |
| Top | 21.40 | 4.87 | 70 |

The resulting photocurrent density and the associated efficiency enhancement for the different vertical position cases shown in Fig. 5 are listed in Table 2. As can be seen there, the optimum vertical position is the "top" one, while the decrease of efficiency going from top to bottoms is not monotonic.

To reveal the origin of this puzzling non-monotonic behavior as we change the vertical position of the metal nanoparticles inside the perovskite layer, we examined for all cases, including that of the pristine PSC, the electric field intensity, $|E|^2/|E_0|^2$. This intensity for $\lambda=673$ nm is plotted in Fig. 6. For the case without nanoparticles, shown in Fig. 6(a), we see that the electric field forms a standing wave inside the perovskite layer which arises from a fabry-perrot resonance induced by the aluminum back reflector of the device (see Fig. S2). Analyzing the plots shown in Fig. 6, we see that we have larger enhancement in the cases where the vertical nanoparticle position coincides with field maxima of the standing wave, and minimum or no enhancement where the sphere position is in the nodes of the electric field. This is easy to understand taking into account that the local field (one of the main factors of the absorption enhancement) is directly proportional to the input field. Thus, nanoparticles at the field maxima experience the largest possible input field associated with the largest resulting local field.

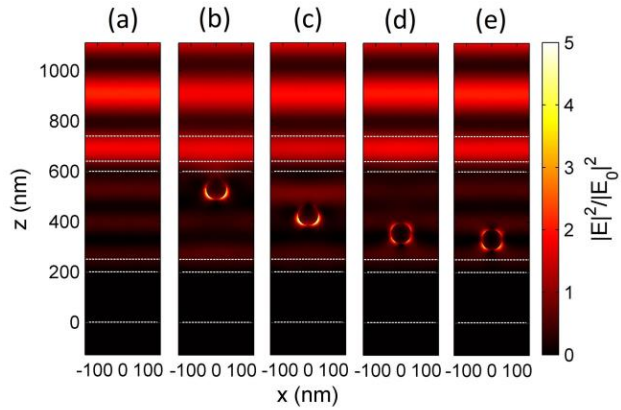

Fig. 6. Normalized distribution of the squared amplitude of the electric field, $|E|^2/|E_0|^2$, at $\lambda=673$nm, for the pristine PSC (a), and for a PSC with spherical silver nanoparticles placed at the four different vertical positions shown in Fig. 5, i.e. at the top position (b), middle (c), bottom (d), and at the bottoms (e). In all cases the nanoparticles radius is 40 nm and their lateral spacing 300 nm. $E_0$ is the incident electric field.

We noticed the same vertical position dependence when spheres of larger size, i.e. with a radius equal to 70 nm, are placed inside the perovskite layer. Although larger spheres are associated with smaller ohmic losses and larger near-field enhancement (as they exploit also the quadrupole besides dipole mode), they do not always offer larger absorption enhancement in PSCs since the additional light trapping offered by the larger spheres at specific wavelengths may be counteracted by the larger amount of the absorbing perovskite material which they replace. Indeed, this is the case in our PSC system, where the absorption enhancement is favored in most of the vertical positions from spheres of radius 40 nm, while larger spheres lead to smaller enhancement.

An exception is observed for spheres placed close to the top of the perovskite layer; there spheres of radius 70 nm gave larger absorption enhancement compared to the 40 nm ones (see last line of Table 2). The basic reason is that spheres close to the top of the perovskite behave also as antireflection layer, directing the scattered light towards perovskite and preventing it from escaping to the top layers. Close to the top of the perovskite layer the larger scattering and field concentration which is offered by the larger size particles leads to larger enhancement.

We note though that if placing large metal spheres at the top positions in perovskite, close to the interface with PEDOT:PSS, there is the risk of increasing the interface transport resistance and therefore a coating layer causing electrical isolation may be needed[28].

### 3.2.3. Nanoparticles concentration effect on the PSC performance

In the current section we aim to elucidate the effect of the concentration of metal particles at the expense of the perovskite material. Concentration is considered a critical parameter in plasmonic solar cells since the precise control of position and size of nanoparticles is not possible at processing, resulting in a more randomized environment. Moreover, to predict the effect of the concentration on the solar cell performance and to optimize the concentration is not a straightforward task as this effect is not expected to be monotonic. Small sphere concentrations are expected to dilute the sample causing a decline of the plasmonic impact while larger concentrations come at the expense of the highly absorbing perovskite material, and are associated with inter-particle coupling, i.e. modified scattering response the effect of which is not straightforward to be predicted, as well as higher ohmic losses. Here we omit the case of very-closely spaced nanoparticles which is associated with the appearance on new inter-particle modes.

In our numerical study, we change the nanoparticles concentration by changing the lattice constant, i.e., the interparticle distance, maintaining a constant sphere radius ($r$=70 nm) and vertical position ("top" position – see Fig. 5), values/conditions which were found as optimum, according to the analysis of our previous sections. The results are shown in Fig. 7, where we plot the absorption versus wavelength for different lattice constants [Fig. 7(a)] and the resulting photocurrent density for the different concentrations investigated [Fig. 7(b)].

As was expected and mentioned above the effect of concentration on the photocurrent density is not monotonic. Here lattice constants around 300 nm seem to give the optimum performance, while both smaller and larger lattice constants lead to a decline of the achieved photocurrent. The achievement of optimum performance for $L$=300 nm, where no strong interparticle coupling is expected (due to the relatively large interparticle distance), indicates that the effects of deviations from the periodicity of the particles system are expected to be relatively small. This is indeed verified by additional simulations (not presented here), where we introduced deviations from the periodicity in the lateral position of the spheres, and it is particularly important since, as was already mentioned, in realistic solar cells the precise control of the position of nanoparticles is not possible at processing, resulting to a rather random plasmonic system.

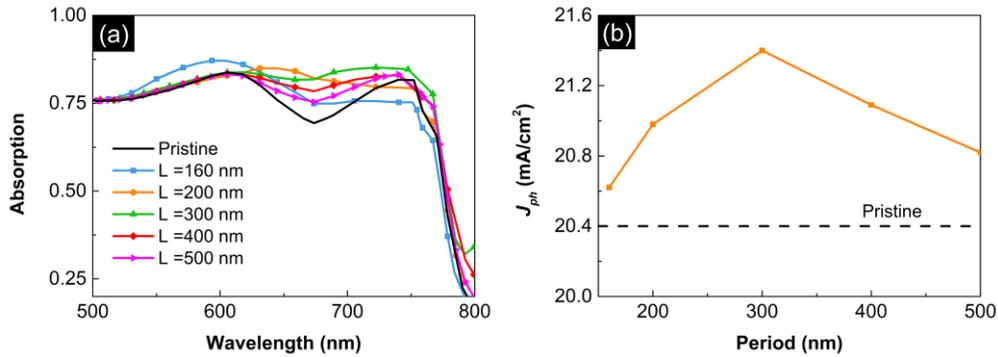

Fig. 7. (a) Absorption of perovskite versus wavelength for different lattice constants (160 nm – blue line; 200 nm – orange line; 300 nm – green line; 400 nm – red line; and 500 nm – magenta line) compared to the pristine case (black line) and (b) the resulting photocurrent density for the different concentrations investigated, assuming silver spheres with a constant radius of 70 nm, and vertical position the "top" position shown in Fig. 5.

### *3.3. Plasmonic nanoparticles inside different than the perovskite layers of PSC*

Due to the manufacturing procedure of the inverted PSC, the embedding of solution-processed metallic nanoparticles at different layers of the device other than the perovskite, e.g. inside the PEDOT:PSS and the PCBM carrier transporting layers, is possible without increasing the cost[16] as well as maintaining the planar architecture[29]. This way, plasmonic enhancement could be exploited without reducing the perovskite material, albeit at the price of reduced local field enhancement in the absorbing perovskite layer.

Furthermore, placing nanoparticles at the carrier transporting layers may facilitate other internal processes related to the electrical properties of the device. For instance, resonant metallic nanoparticles have been shown to be able to favor the electrical properties of the system too[16],[29],[30]. Increased exciton generation and dissociation (photocarrier generation, separation) due to the localized surface plasmon resonance oscillating fields that also extend inside the perovskite layer[12], improved carrier transport and extraction due to lower series or contact resistance[16],[30] of the solar cell and stability[31] are some examples of the additional impact of plasmonics in PSCs.

In this work we aim first to examine and improve the optical absorption of the PSC by incorporating metal nanoparticles inside the hole-transporting layer, PEDOT:PSS; this way we hope to minimize the reflection at the PEDOT:PSS-Perovskite interface. In the previous discussion we have revealed the importance of the reflection minimization, especially at the PEDOT:PSS-Perovskite interface. Particularly, we showed that the reflection from the multilayered device is the main reason resulting in decreased absorption in perovskite [Fig. 2(b)] along the entire spectrum (300<$\lambda$<800 nm), especially for $\lambda$<550 nm. Interestingly, the low refractive index of PEDOT:PSS ($n$~1.41) tunes the plasmonic resonances of metal nanoparticles at lower wavelengths (compared to the case of perovskite host) which coincide with this spectral region. The main restriction in the introduction of nanoparticles in the HTL/ETL is the small thickness of those layers (40, 50 nm) that limits the size of nanoparticles able to maintain the simplicity of the planar architecture. Employing small spheres, though, results to increase of the ohmic losses, as for small spheres absorption dominates extinction. Indeed, embedding spheres of radius $r$~18 nm in the PEDOT:PSS enhancement was achieved only for the case of aluminum, with maximum achievable photocurrent $J_{ph}$ =21.22 mA/cm$^2$ [see Fig. 8(a)], corresponding to an enhancement of ~4% (assuming a perovskite layer thickness of 350 nm, and a spheres period of 65 nm). The photocurrent enhancement by embedding the nanoparticles in the PEDOT:PSS layer indicates the possibility to reduce the thickness of the perovskite layer, reducing thus the toxicity of the device (due to the lead content reduction), maintain though its efficiency. Indeed, as is verified by related simulations, calculating the photocurrent for different perovskite layer thicknesses [see Fig. 8(b)], we observe that thicknesses of about ~270 nm generates the same $J_{ph}$ with that of the conventional (pristine) PSCs with the optimum thickness of ~350 nm.

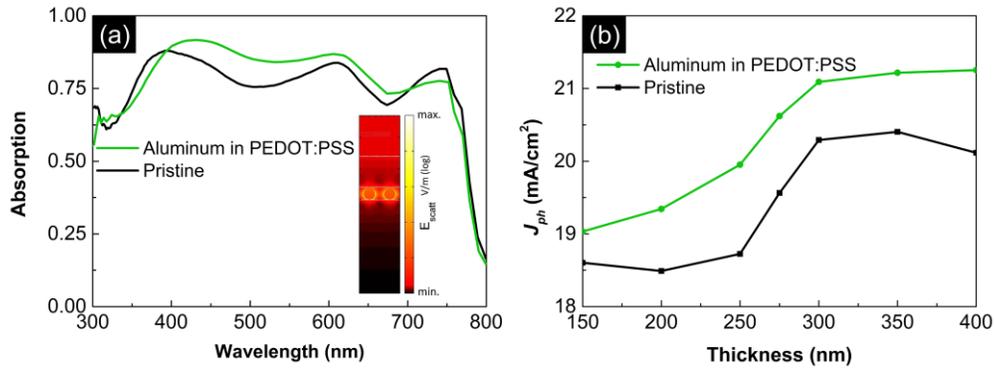

Fig. 8. (a) Absorption in perovskite for the pristine (black line) and the plasmonic PSC (green line) assuming aluminum spheres with radius equal to 18 nm and a period of 65 nm placed inside

the PEDOT:PSS carrier transporting layer. The origin of the absorption enhancement is depicted at the right inset where the scattered field, due to the presence of the aluminum nanoparticles, is plotted for $\lambda$=517 nm. (b) $J_{ph}$ of the pristine (black line), plasmonic PSC (green line) as a function of the perovskite thickness.

The origin of the absorption enhancement in the case of nanoparticles embedded in the PEDOT:PSS is the re-distribution of the incident light with enhanced scattering in the forward direction (towards the perovskite layer) along with the high local fields which are also extended inside the perovskite layer [see inset in Fig. 8(a)], in conjunction with the low parasitic absorption in aluminum nanoparticles.

Summarizing, we have to note that the antireflection property in PSCs with plasmonic nanoparticles placed in PEDOT:PSS is a quite prominent absorption enhancement approach especially at lower wavelengths (<550 nm) because there is no competition between the volume occupied by the metal particles and the absorptive perovskite material. Employing this approach allows employment of thinner perovskite layers, and thus reduction of the structure toxicity, without scarifying the PSC performance. Placing nanoparticles in PEDOT:PSS is considered a reliable strategy regarding the enhancement of the overall efficiency of the solar cell since studies have been shown it to favor the electrical properties of the system too[16],[29].

Regarding incorporation of nanospheres inside the electrons transporting layer (PCBM), we found no enhancement, given the already known nanoparticle size restrictions owing to the small thickness of the ETL as well as the impossible exploitation of antireflection at lower wavelengths (where the plasmonic resonance of nanospheres in PCBM, with $n$~2, occurs) in such depth of the PSC, given the highly absorptive nature of perovskite.

An approach to utilize the higher wavelengths (>500 nm) to achieve an efficiency enhancement is to utilize nanoparticles of different shape or different aspect ratio[32] to tailor the dispersion properties of their localized surface plasmon resonances and tune them at higher wavelengths [Fig. S4(a)]. The most suitable candidates seem to be the nanorod-shaped particles, because they are synthesized in a wide range of aspect ratios[33] showing two localized dipole resonant modes (aligned with their short and long axis) and the mode aligned with the axis parallel to the PCBM layer (where no significant size restrictions exist) is highly tunable. The incorporation though of nanorod particles in the ETL layer of our system showed very limited absorption enhancement [Fig. S4(b)] (due to increased near-field intensity at the vicinity of the nanorods that also extended inside the perovskite material [see inset in Fig. S4(b)]), which was polarization dependent and it is questionable if can be observed in realistic systems (where disorder in the nanoparticle parameters is unavoidable).

Therefore, we conclude that utilizing plasmonic nanoparticles inside the PCBM is not recommended for improving the optical response of the PSCs. Other techniques should be exploited here, i.e., nanostructuring on the back-reflector[34] as long as the interface transport resistance does not increase. Lastly, we note that nanoparticles can be placed inside the ITO layer too with a less harsh size constriction (ITO thickness ~100 nm). However, there they cannot contribute to the increase of the local fields inside the perovskite (since they are far from perovskite), thus their basic role will be to act as antireflectors. For that, nanoparticles of more complex geometry[35] or different shape[29] seem to be more appropriate than the spherical ones.

*3.4. Optimizing PSC performance by combining plasmonic particles in different PSC layers*

In this last section of our paper we aim to exploit the full potential of plasmonics in PSCs following our conclusions at earlier parts. This way, an enhanced absorption efficiency, compared to the conventional PSCs (with optimum thicknesses of about ~350 nm), could be achieved as well as the amount of the lead could be further reduced, using thinner perovskite absorbers or replacing perovskite by the nanoparticle material, without deteriorating the absorption.

The first step to explore the possibility of PSCs with reduced lead and high absorption efficiency is to examine and exploit the additive-like absorption behavior of spectrally separated resonances originating from nanoparticles at different PSC layers. For that reason, we utilize the combined effect of both aluminum nanospheres placed inside the PEDOT:PSS layer and of silver spheres located inside the perovskite. Aluminum spheres in PEDOT:PSS provide absorption enhancement at the range of 400<$\lambda$<700 nm while silver spheres of optimized size, spacing and position, placed inside the perovskite layer, provide absorption enhancement at the region of 650<$\lambda$<800 nm.

Fig. 9(a) shows the absorption in perovskite of the combined case compared to the "individual" cases assuming a perovskite layer thickness of ~350 nm. Interestingly, a definite "additive" response is depicted, which is preserved for the whole range of perovskite layer thicknesses, from 400 up to 150 nm [see Fig. 9(b)], confirming the robustness of this approach. Indeed, the enhancement of the combined case equals the sum of the enhancements for only silver (in perovskite) and only aluminum (in PEDOT:PSS) spheres. We have to note here that the parameters used in this calculation have been obtained from the optimization studies discussed in the previous sections, with minor modifications as to accommodate simulations, to minimize the probability to affect the electrical properties of the PSC and to ensure optimized combined case (i.e. with spheres in both PEDOT:PSS and perovskite layers).

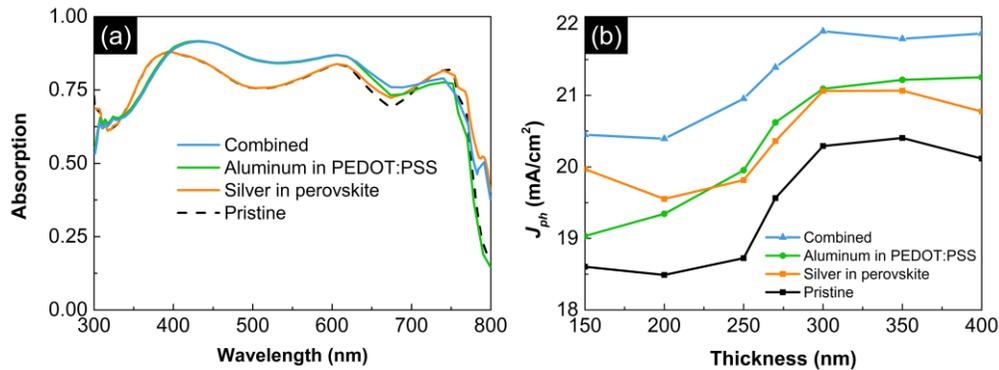

Fig. 9. (a) Absorption in perovskite with thickness equal to 350 nm for the following cases: pristine device (black dashed line), aluminum spheres placed inside the PEDOT:PSS (green line – sphere radius 18 nm, period 65 nm), silver spheres inside the perovskite in the position middle (orange line – sphere radius 30 nm, period 325 nm) and their combined case (blue line). (b) Jph as a function of the perovskite thickness for the plasmonic PSCs. The green line corresponds to the case when only aluminum spheres (radius 18 nm, period 65 nm) are placed inside the PEDOT:PSS, the orange line corresponds to the case when only silver spheres are placed inside the perovskite with optimized vertical position, radius and concentration, the blue line shows the effect of the combination of aluminum spheres inside the PEDOT:PSS and silver spheres inside the perovskite; all results are compared to the pristine case (black line).

From Fig. 9(b) one can see that for a perovskite layer thickness equal to 150 nm the photocurrent reaches the record value $J_{ph}$ = 20.45 mA/cm$^2$. This result indicates that the lead content reduction can reach values up to 43% (taking into account only the thickness reduction), reducing this way the toxicity of the PSC by a great amount without reducing the photocurrent efficiency compared to conventional pristine PSCs with optimum perovskite thicknesses of about ~350 nm. Regarding the PSC of perovskite thickness 350 nm we found a maximum $J_{ph}$ = 21.80 mA/cm$^2$, corresponding to an enhancement of 6.8% relative to pristine PSC.

We have to note here that the calculated $J_{ph}$ was even higher than the values shown in Fig. 9(b), and equal to the champion 22.11 mA/cm$^2$ (~8.4% increase compared to pristine), assuming larger silver spheres, with $r$=70 nm, placed at the "top" position of the perovskite layer with a thickness equal to 350 nm. The results in Fig. 9(b) though correspond to smaller spheres owing to the risk of increasing the interface transport resistance.

## 4. Conclusions

We examined and discussed here the effect of plasmonic nanoparticles in realistic perovskite solar cells. Placing spherical nanoparticles in different layers of the PSC we examined the possibility to achieve broadband enhancement in the light absorption and thus enhanced photocurrent density, as well as the related conditions regarding nanoparticle material, size, vertical position and concentration, in particular for nanoparticles embedded in the perovskite layer. Our study showed optimum response for silver nanoparticles of radius close to 40 nm if placed in the middle or at the top of the perovskite layer and of even larger radius (e.g. 70 nm) if placed close to the top of the layer, while the optimum nanoparticle distance was found to be around 300 nm. For silver spheres in the middle of the layer the maximum photocurrent density enhancement was found equal to 3.24% corresponding to a $J_{ph}$ equal to 21.07 mA/cm$^2$ while for silver spheres close to the top of the perovskite layer the photocurrent enhancement reached values of 4.87% corresponding to a $J_{ph}$ of 21.40 mA/cm$^2$. The origin of the photocurrent enhancement in all cases was the high local field values associated with the plasmonic resonances (which maximized the absorption), together with the enhanced scattering and antireflection properties of those particles, especially if placed close to the top of the solar cell. The calculated enhancement was found to be quite robust against nanoparticles randomness in the lateral position and polydispersity.

We found considerable photocurrent enhancement (up to 4.0% corresponding to $J_{ph}$=21.22 mA/cm$^2$) also for nanoparticles placed in the hole transporting layer (PEDOT:PSS, on top of the perovskite layer with a thickness 350 nm) despite the small thickness of the HTL layer and the associated restrictions of nanoparticle size. The origin of the photocurrent enhancement was the enhanced scattering in the forward direction (towards the perovskite layer) along with the high local fields which are also extended inside the perovskite layer. Here aluminum nanospheres (of $r$=18 nm and distance ~65 nm) gave the optimum performance.

The combined effect of placing both aluminum nanospheres in PEDOT:PSS and silver spheres in perovskite resulted at the champion 8.36% enhancement, corresponding to $J_{ph}$ equal to 22.11 mA/cm$^2$, confirming the additive like absorption behavior of spectrally separated resonances originating from nanoparticles placed at different positions of the solar cell. The absorption enhancement offered by the plasmonic nanoparticles indicates the potential to employ PSCs with much thinner perovskite layers, e.g. 150 nm (with reduced toxicity due to the reduced amount of lead), maintaining though the performance of the current PSCs of optimum thickness 350 nm.

Summarizing, we found that embedding metal nanospheres inside realistic PSCs architectures provides a great strategy to improve the optical performance of the device as well as to decrease its toxicity without increasing the cost or the fabrication complexity.

## Funding

This research has been co-financed by the European Union and Greek national funds through the Operational Program Competitiveness, Entrepreneurship and Innovation, under the call RESEARCH – CREATE – INNOVATE (T1EDK-01082).

# Supporting Information

# Efficient and environmental-friendly perovskite solar cells via embedding plasmonic nanoparticles: an optical simulation study on realistic device architecture

George Perrakis,[*,1,2] George Kakavelakis,[*,3,4] George Kenanakis,[1] Constantinos Petridis,[3,5] Emmanuel Stratakis,[1] Maria Kafesaki,[1,2] and Emmanuel Kymakis[3]

[1]Institute of Electronic Structure and Laser (IESL), Foundation for Research and Technology Hellas (FORTH), Heraklion, Greece
[2]Dept. of Materials Science and Technology, Univ. of Crete, Heraklion, Greece
[3]Center of Materials Technology and Photonics and Electrical Engineering Department, Technological Educational Institute (TEI) of Crete, Heraklion, Crete, 71004, Greece
[4]Cambridge Graphene Centre, University of Cambridge, 9 JJ Thomson Avenue, Cambridge CB3 0FA, UK
[5]Department of Electronic Engineering Technological Educational Institute (TEI) of Crete, Chania 73132, Greece
Corresponding authors:
*E-mail: gperrakis@iesl.forth.gr
*E-mail: gk415@cam.ac.uk

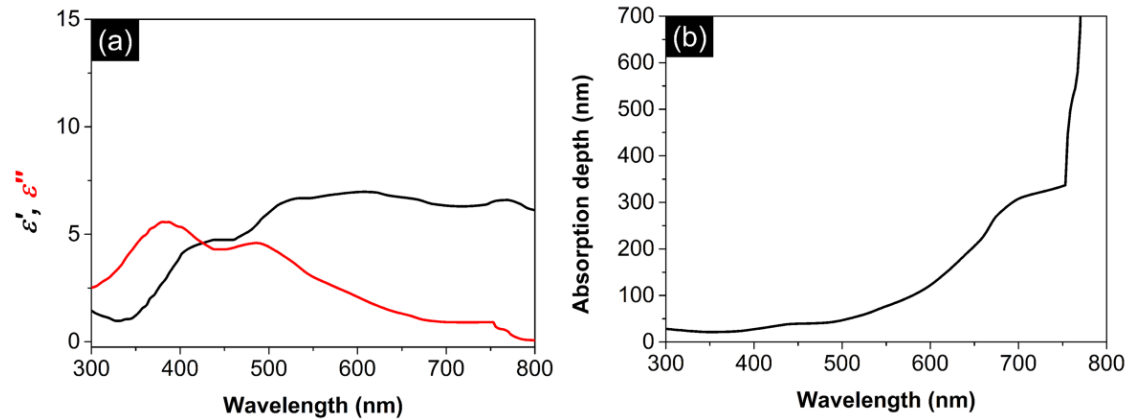

Fig. S1. (a) Real ($\varepsilon'$ – black line) and imaginary ($\varepsilon''$ – red line) parts of the perovskite permittivity considered in the calculations. (b) Absorption depth in perovskite given by the inverse of the absorption coefficient.

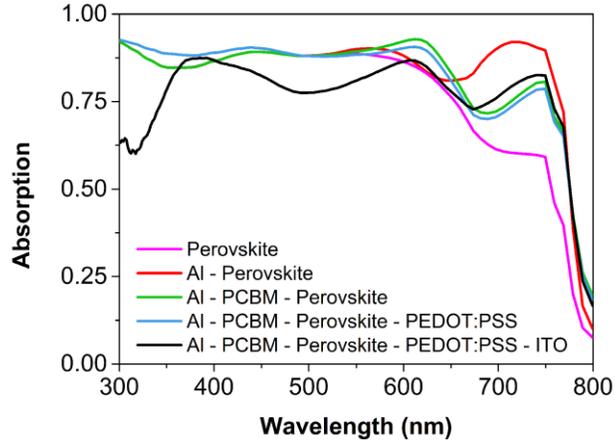

Fig. S2. Interference effects introduced by the several layers of the device of finite thickness.

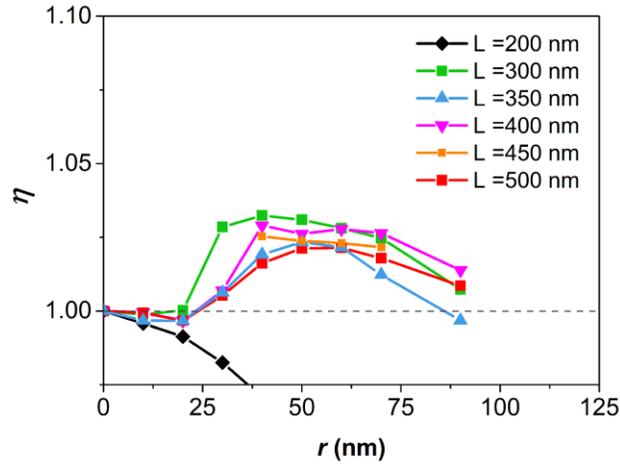

Fig. S3. Optimization process for silver nanoparticles, where we required maximum photocurrent density enhancement, $\eta$, for the plasmonic PSC relative to the pristine PSC, for spheres located exactly at the middle of the perovskite layer (along $z$-direction – see Fig. 1) by changing the radius, $r$, of the spheres, and by changing the periodicity, $L$ ($L = L_x = L_y$), at which spheres are placed along the $x$- and $y$-axis.

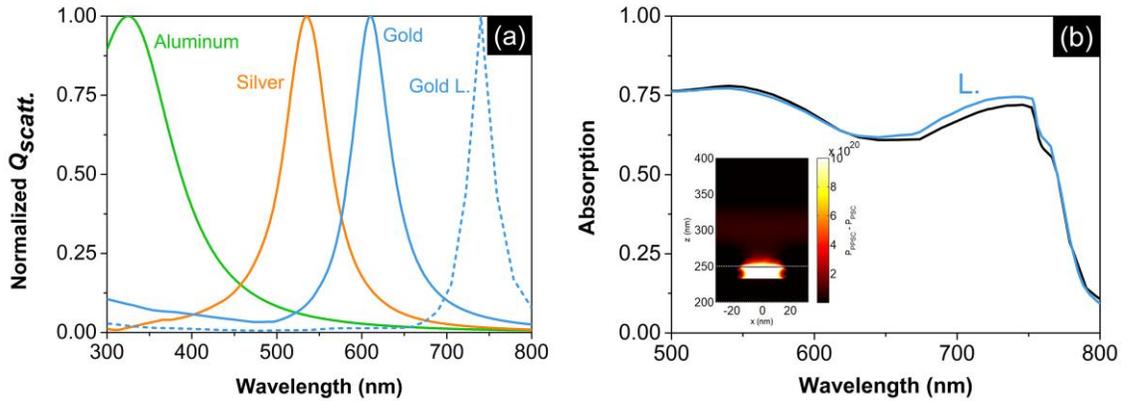

Fig. S4. (a) The utilization of gold nanorods tunes the plasmonic resonances ($Q_{scatt.}$: scattering cross-section) at higher wavelengths compared to spherical nanoparticles. (b) Absorption enhancement relative to the pristine case (black line) due to the excitation of the longitudinal plasmonic mode of the gold nanorods placed inside the PCBM. The origin of the enhancement is depicted at the right inset where the power (PPPSC - PPSC in W/m$^3$) concentrated around the gold nanorods that extends inside the perovskite layer is plotted.